\documentclass[twocolumn,showpacs,preprintnumbers,amsmath,amssymb,superscriptaddress]{revtex4}

\usepackage{graphicx,dcolumn,bm,mathrsfs}

\newcommand{\nl}{\nonumber \\}
\newcommand{\be}{\begin{equation}}
\newcommand{\ee}{\end{equation}}
\newcommand{\bea}{\begin{eqnarray}}
\newcommand{\eea}{\end{eqnarray}}
\newcommand{\bsube}{\begin{subequations}}
\newcommand{\esube}{\end{subequations}}

\newcommand{\Eq}[1]{Eq.\,(\ref{#1})}

\newcommand{\Fig}[1]{Fig.\,\ref{#1}}

\newcommand{\rmL}{{\rm L}}
\newcommand{\rmR}{{\rm R}}
\newcommand{\rmc}{{\rm c}}
\newcommand{\rmi}{{\rm i}}
\newcommand{\rmd}{{\rm d}}
\newcommand{\ra}{\rangle}
\newcommand{\la}{\langle}
\newcommand{\alf}{\alpha}
\newcommand{\sgm}{\sigma}
\newcommand{\Omg}{\Omega}
\newcommand{\omg}{\omega}
\newcommand{\Gam}{\Gamma}
\newcommand{\Dlt}{\Delta}
\newcommand{\dlt}{\delta}
\newcommand{\vpl}{\varepsilon}
\newcommand{\epl}{\epsilon}

\newcommand{\Lmd}{\Lambda}
\newcommand{\GamL}{\Gamma_{\rm L}}
\newcommand{\GamR}{\Gamma_{\rm R}}
\newcommand{\NL}{{N_{\rm L}}}
\newcommand{\NR}{{N_{\rm R}}}
\newcommand{\tr}{{\rm tr}}

\begin{document}


\title{Unraveling of a generalized quantum Markovian master equation and its application in
feedback control of a charge qubit}

\author{JunYan Luo}\email{jyluo@zust.edu.cn}
\affiliation{Department of Physics, Zhejiang University of Science
  and Technology, Hangzhou 310023, China}
\author{Jinshuang Jin}
\affiliation{Department of Physics, Hangzhou Normal University,
Hangzhou 310036, China}
\author{Shi-Kuan Wang}
\affiliation{Department of Physics, Hangzhou Dianzi University,
Hangzhou 310018, China}
\author{Jing Hu}
\affiliation{Department of Physics, Zhejiang University of Science
  and Technology, Hangzhou 310023, China}
\author{Yixiao Huang}
\affiliation{Department of Physics, Zhejiang University of Science
  and Technology, Hangzhou 310023, China}
\author{Xiao-Ling He}
\affiliation{Department of Physics, Zhejiang University of Science
  and Technology, Hangzhou 310023, China}

 \date{\today}

 \begin{abstract}
 In the context of a charge qubit under continuous monitoring
 by a single electron transistor, we propose an unraveling of
 the generalized quantum Markovian master equation into an ensemble of
 individual quantum trajectories for stochastic point process.
 A suboptimal feedback algorism is implemented into individual 
 quantum trajectories to protect a desired pure state.
 Coherent oscillations of the charge qubit could be maintained
 in principle for an arbitrarily long time in case of sufficient
 feedback strength.
 The effectiveness of the feedback control is also reflected 
 in the detector's noise spectrum.
 The signal-to-noise ratio rises significantly with increasing
 feedback strength such that it could even exceed the
 Korotkov-Averin bound in quantum measurement, manifesting 
 almost ideal quantum coherent oscillations of the qubit.
 The proposed unraveling and feedback protocol may open up
 the prospect to sustain ideal coherent oscillations
 of a charge qubit in quantum computation algorithms.
 \end{abstract}
`



\maketitle

\section{\label{thsec1} Introduction}

 The advent of quantum information technologies is creating a
 considerable demand for strategies to manipulate individual
 quantum systems in the presence of noise \cite{Wis10}.
%
 Recently, state-of-the-art nano-fabrication has made it possible
 to accurately monitor a single quantum state in a continuous
 manner \cite{Fuj061634,Gus09191,Vij11110502}.
 It opens up the prospect for physically implementing the
 measurement-based (close-loop) feedback control of a quantum
 system, in which the real-time information of the detector's
 output is extracted and instantaneously fed appropriate
 corrections back onto it \cite{Vij1277,Say1173,Bra12173601,Gil10080503,Koc10173003}.
 So far, a variety quantum feedback protocols have been
 proposed \cite{Rus02041401,Kor05201305}.
 These control strategies turn out to be efficient for potential
 applications ranging from the realization of a mesoscopic Maxwell's
 demon \cite{Sch11085418,Str13040601,Esp1230003} and purification
 of a charge qubit \cite{Kie11050501,Pol11085302,Kie12123036},
 to the freezing of charge distribution \cite{Bra10060602} in
 full counting statistics (FCS) \cite{Bla001,Naz03}.

 In contrast to the classical feedback, by which one might acquire as
 much system information as possible, in quantum feedback control
 the measured system is inevitably disturbed in an unpredictable
 manner, and the amount of information is fundamentally limited
 by the Heisenberg's uncertainty principle.
 The essence of the involving measurement process is the trade-off
 between the acquisition of information and the backaction induced
 dephasing of the quantum state \cite{Kor995737,Kor01115403,Gur9715215,Mak01357}.
 It is therefore suggested to consider a continuous weak quantum
 measurement, in which information gain and disturbance of the measured
 system could reach a balance.
 The Lindblad master equation is, in general, utilized for continuous
 measurement and quantum feedback control.
 The unique advantage is its clear physical implications and simplicity
 to be unraveled into individual quantum trajectories, in which quantum
 feedback operations might be appropriately implemented \cite{Wis10,Jac14}.

 It was revealed, however, the Lindblad master equation may not 
 satisfy the detailed balance relation \cite{Gor78149}.
 Furthermore, the dynamics of a quantum system does not necessarily obey 
 the Lindblad master equation in some parameter regimes (for 
 instance, the applied voltage is comparable to the internal energy 
 scales of the quantum system).
 A generalized quantum Markovian master equation (G-QMME) has been employed
 to study the unique features involved in various mesoscopic structures,
 such as super-Poissonian noise in transport double quantum dot systems \cite{Wun05205319,Luo11145301,Luo1159},
 additional dephasing in an Aharonov-Bohm interferometer \cite{Mar03195305,Bru96114},
 as well as ``an exchange magnetic field'' in a quantum dot spin
 valve structure \cite{Kon03166602,Bra06075328,Sot10245319}.
 Yet, direct application of the G-QMME to quantum feedback control is
 restricted due to its difficulty to be unraveled into individual
 trajectories.
 An essential issue, therefore, is to find a recipe which is
 capable of unraveling the G-QMME with specific physical meanings
 such that it can be applied in the realm of quantum feedback
 control.

%

\begin{figure*}
\begin{minipage}[c]{.51\textwidth}
\includegraphics[scale=0.47]{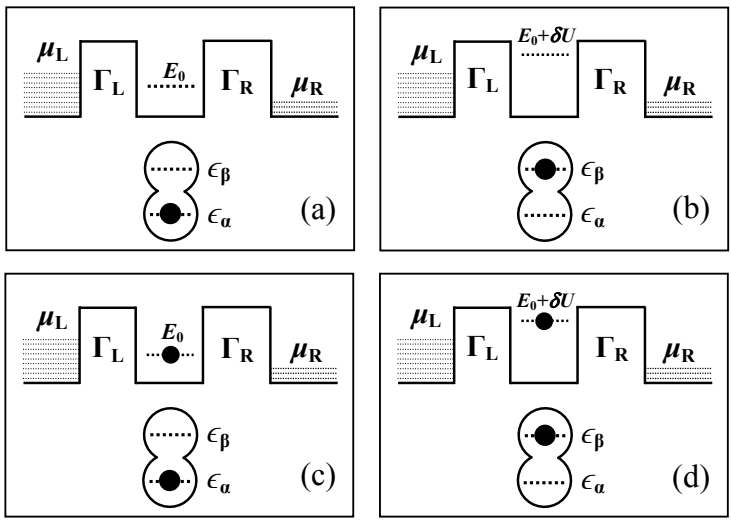}
\end{minipage}
\hspace{0.08\textwidth}
\begin{minipage}[c]{.38\textwidth}
 \caption{\label{Fig1}Schematic setup of a solid-state charge
 qubit under continuous monitoring by an SET detector.
 The charge qubit is represented as an extra electron confined
 in a double quantum dot.
 The SET detector is a single QD sandwiched between the left
 and right electrodes, where the quantum dot is capacitively
 coupled to the qubit via Coulomb repulsion.
 Possible electron configurations of the reduced quantum system (qubit
 plus SET QD) are displayed in (a)-(d), respectively.}
\end{minipage}
\end{figure*}

 In this work, we propose a physical unraveling of the G-QMME
 in the context of qubit measurement by a single electron
 transistor (SET).
 It is then implemented to stabilize coherent evolution of
 the measured qubit based on a suboptimal feedback
 algorithm \cite{Doh01062306,Jin06233302}.
 Actually, an unraveling scheme of a G-QMME was recently
 constructed for a diffusion process \cite{Wan07155304},
 which, however, limited its applications only to diffusive
 detectors, like a quantum point contact (QPC).
 For an SET detector, the measurement current exhibits a typical
 stochastic point process, which comprises unique advantages of high
 sensitivity and triggering feedback operations straightforwardly.
 An unraveling scheme tailored to this process is therefore of
 crucially importance for physically implement the measurement-based
 feedback control in the context of an SET detector.

 The remainder of the paper is organized as follows.
 We start with a description of the measurement setup in section
 \ref{thsec2}.
 The G-QMME for the dynamics
 of the reduced system is presented in section \ref{thsec3}.
 It is then followed in section \ref{thsec4} by the unraveling
 of the G-QMME into an ensemble
 of individual  quantum trajectories, in which the dynamics of
 the reduced  system in every single realization of continuous
 measurements is conditioned on stochastic point process.
 Quantum trajectories and corresponding stochastic tunneling
 events are numerically simulated in section \ref{thsec5} based
 on the Monte Carlo scheme.
 Section \ref{thsec6} is devoted to the implementation of feedback
 control protocol to protect coherent oscillation of the qubit states,
 which leads to the violation of the Korotkov-Averin bound for a
 sufficient feedback strength.
 Finally, we conclude in section \ref{thsec7}.

 \section{\label{thsec2}Model Description}

 The setup of charge qubit under continuous measurement by an SET detector
 is schematically shown in \Fig{Fig1}.
 The charge qubit is represented as an extra electron in a double
 quantum dot. Whenever the electron occupies the lower (upper) quantum
 dot [cf. \Fig{Fig1}], the qubit is said to be in the logical state
 $|\alf\ra$ ($|\beta\ra$).
 The measurement SET is a device with a single quantum dot (QD) sandwiched
 between the left and right electrodes, where the QD is
 capacitively coupled to the qubit via Coulomb interaction.
 The energy level of the single QD is susceptible to changes
 in the nearby electrostatic environment, and thus can be used to
 sense the location of the extra electron in qubit.
 The Hamiltonian of the entire system reads
 \be\label{Htot}
 H=H_{\rm S}+H_{\rm B}+H'.
 \ee
 The first part
 \be\label{Hs}
 H_{\rm S}=\frac{1}{2}\epl\, \sgm_z + \Omg\, \sgm_x+(E_0+\dlt U)d^\dag d,
 \ee
 describes the charge qubit, SET QD and their coupling,
 where $d^\dag$ ($d$) is the creation (annihilation) operator
 for an electron in the QD, $\epl$ is the qubit level mismatch,
 $\Omg$ is the qubit interdot tunneling, and
 $\sgm_z\equiv|\alf\ra\la \alf|-|\beta\ra\la \beta|$ and
 $\sgm_x\equiv|\alf\ra\la \beta|+|\alf\ra\la \beta|$ are pseudo-spin
 operators for the qubit.
 We assume there is only one energy level $E_0$ within the bias
 regime defined by the Fermi energies in the left and right
 electrodes of the SET.
 Whenever the qubit occupies the logical state $|\beta\ra$, the
 SET QD ``feels'' the existence of a strong Coulomb
 repulsion $\dlt U$ (as inferred in \Fig{Fig1}), which influences
 transport through the SET.
 It is right this mechanism that the qubit information may be
 acquired via the readout of the SET.

 The left and right electrodes of the SET are modeled as reservoirs
 of noninteracting electrons, with Hamiltonian
 \be
 H_{\rm B}=\sum_{\ell={\rm L,R}}\sum_k\vpl_{\ell k}c^\dag_{\ell k}c_{\ell k}.
 \ee
 Here $c^\dag_{\ell k}$ ($c_{\ell k}$) stands for creation (annihilation)
 of an electron in the left ($\ell$=L) or right ($\ell$=R) electrode with
 momentum $k$.
 The reservoirs are characterized by the Fermi functions $f_{\rm L/R}(\omg)$,
 in which the involving Fermi energies determine the applied bias voltage,
 i.e. $V=\mu_{\rm L}-\mu_{\rm R}$.
 Throughout this work, we set $\hbar=e=1$ for the
 Planck constant and electron charge, unless stated otherwise.

 The last component of \Eq{Htot} depicts the electron tunneling
 through the SET
 \be
 H'=\sum_{\ell,k} (t_{\ell k}c^\dag_{\ell k}d+{\rm h.c.})
 \equiv\sum_{\ell}(F^\dag_{\ell}d+{\rm h.c.}),
 \ee
 where $F_{\ell}$ and $F^\dag_{\ell}$ are implicitly defined.
 The tunnel-coupling strength to the electrodes is given by the
 intrinsic tunneling width
 $\Gam_{\ell}(\omg)=2\pi\sum_k|t_{\ell k}|^2\dlt(\omg-\vpl_{\ell k})$.
 In what follows we assume wide-band limit, which leads to energy
 independent tunnel-couplings $\Gam_{\rm L/R}$.

\section{\label{thsec3}Generalized Quantum Markovian Master Equation}

 Under the second-order Born-Markov approximation, the dynamics
 of the reduced system (qubit plus SET QD) can be described by the
 following G-QMME \cite{Yan982721,Yan05187,Bre991633,%
 Bra06026805,Fli08150601,Zwa01,Kon961715}
 \bsube\label{QME}
 \begin{gather}
 \dot{\rho}(t)=-\rmi{\cal L}_0\rho(t)-{\cal R}\rho(t),
 \\
 {\cal R}\rho(t)=\frac{1}{2}\{[d^\dag, A^{(-)}\rho(t)
 +\rho(t) A^{(+)}]\}+{\rm h.c.},
 \end{gather}
 \esube
 where ${\cal L}_0(\cdots)\equiv[H_{\rm S},(\cdots)]$ denotes the
 Liouvillian associated with the system Hamiltonian \Eq{Hs}, and
 $A^{(\pm)}=\sum_{\ell={\rm L,R}}A^{(\pm)}_\ell$, with
 $A^{(\pm)}_\ell\equiv[C^{(\pm)}_\ell(\pm{\cal L})+\rmi D^{(\pm)}_\ell
 (\pm{\cal L})]d$.
 The involving bath spectral functions $C^{(\pm)}_\ell(\pm{\cal L})$
 are related to electron tunneling through the left and right junctions
 of the SET detector
 \be
 C^{(\pm)}_\ell(\pm{\cal L})=
 \int_{-\infty}^\infty \rmd t \,C^{(\pm)}_\ell(t) e^{\pm \rmi {\cal L}t}.
 \ee
 Here the bath correlation functions are given by
 \bsube
 \bea
 C_\ell^{(+)}(t)=\la F_\ell^\dag(t)F_\ell\ra_{\rm B},
 \\
 C_\ell^{(-)}(t)=\la F_\ell(t)F^\dag_\ell\ra_{\rm B},
 \eea
 \esube
 with
 $\la \cdots\ra_{\rm B}\equiv{\rm tr}_{\rm B}[(\cdots)\rho_{\rm B}]$, and
 $\rho_{\rm B}$  the local  thermal equilibrium state of the  SET
 electrodes.
 With the knowledge of the bath spectral functions, the dispersion functions
 $D_\ell^{(\pm)}(\pm{\cal L})$ can be evaluated via the Kramers--Kronig
 relation
 \be\label{disp}
 D_\ell^{(\pm)}(\pm{\cal L})=-\frac{1}{\pi}{\cal P}
 \int_{-\infty}^{\infty}\rmd\omg
 \frac{C_\ell^{(\pm)}(\pm\omg)}{{\cal L}-\omg},
 \ee
 where ${\cal P}$ stands for the Cauchy principal value. The dispersion 
 functions are responsible for the renormalization of the internal energy 
 scales, which might have important roles to play in mesoscopic transport. 
 Actually, a number of unique feature have already been
 revealed due to the dispersion functions.
 For instance, a bias dependent phase shift and additional dephasing of 
 the quantum states in a mesoscopic Aharonov-Bohm interferometer is 
 revealed \cite{Bru96114,Mar03195305}.
 In a QD spin valve, the dispersion gives rise to an effective
 exchange magnetic field, which causes prominent spin
 precession behavior \cite{Kon03166602,Bra06075328,Sot10245319}.
 Even in a double quantum dot transport system, the dispersion leads to
 bunching of tunneling events, and eventually to the strong
 super-Poissonian shot noise \cite{Luo11145301,Luo1159}.

 Owing to the presence of the dispersion functions, the G-QMME as shown in
 \Eq{QME} is apparent a {\it non}-Lindblad-type master equation.
 How to unravel the G-QMME in an appropriate way such that the quantum
 feedback may be implemented in every individual quantum trajectories
 become an essential issue in quantum control.
 In the context of continuous qubit measurement by an SET detector, we
 will present a physical unraveling of the G-QMME [\Eq{QME}] exclusively 
 for the point process in the following section.

\section{\label{thsec4}Unraveling of the G-QMME}

 The G-QMME [\Eq{QME}] only depicts the dynamics of the reduced system
 under continuous measurement by an SET detector. In order to achieve
 a description of the readout from the SET detector, we shall resolve
 the number of electrons transport through the SET in \Eq{QME}, and
 arrive at a number-resolved quantum master equation \cite{Li05205304,Luo07085325}
 \be\label{CQME}
 \dot{\rho}^{(\NL,\NR)}=-\left\{\rmi{\cal L}_0+{\cal R}_0+{\cal R}^{\pm}_{\rm L}
 +{\cal R}^{\pm}_{\rm R}\right\}\rho^{(\NL,\NR)},
 \ee
 where
 \bsube
 \be
 {\cal R}_0\rho^{(\NL,\NR)}=\frac{1}{2}\{d^\dag A^{(-)}\rho^{(\NL,\NR)}+\rho^{(\NL,\NR)} A^{(+)}d^\dag\}+{\rm h.c.}
 \ee
 represents continuous evolution of the reduced system, while
 \begin{gather}
 {\cal R}^{+}_{\rm L}\rho^{(\NL,\NR)}
 =-\frac{1}{2}\{d^\dag \rho^{(\NL+1,\NR)}A^{(+)}_\rmL\}+{\rm h.c.}
 \\
 {\cal R}^{-}_{\rm L}\rho^{(\NL,\NR)}
 =-\frac{1}{2}\{A^{(-)}_\rmL \rho^{(\NL-1,\NR)}d^\dag\}+{\rm h.c.}
 \end{gather}
 and
 \begin{gather}
 {\cal R}^{+}_{\rm R}\rho^{(\NL,\NR)}
 =-\frac{1}{2}\{d^\dag \rho^{(\NL,\NR+1)}A^{(+)}_\rmR\} +{\rm h.c.}
 \\
 {\cal R}^{-}_{\rm R}\rho^{(\NL,\NR)}
 =-\frac{1}{2}\{A^{(-)}_\rmR \rho^{(\NL,\NR-1)}d^\dag\} +{\rm h.c.}
 \end{gather}
 \esube
 describes electron transport through the left and right junctions
 of the SET detector.
 Here $N_{\ell}$ denotes the number of electrons tunneled through
 the left ($\ell$=L) or right ($\ell$=R) junction of the SET up to
 time $t$.
 This equation cannot be solved directly, since it actually corresponds to
 an infinite number of coupled equations.
 We thus perform a discrete Fourier transformation
 $\tilde{\rho}(\chi_\rmL,\chi_\rmR,t)=\sum_{\NL,\NR}
 e^{\rmi (\NL \chi_\rmL+\NR \chi_\rmR)} \rho^{(\NL,\NR)}(t)$
 and finally arrives at a $\chi$-resolved master equation
 \begin{align}\label{vrho}
 \frac{\partial}{\partial t}\tilde{\rho}
 &=-\left\{\rmi{\cal L}_0+{\cal R}_0+\sum_{\pm}\left(e^{\mp\rmi\chi_\rmL}{\cal R}^{\pm}_{\rm L}
 +e^{\pm\rmi\chi_\rmR}{\cal R}^{\pm}_{\rm R}\right)\right\}\tilde{\rho}
 \nonumber
 \\
 &\equiv{\cal L}(\chi_\rmL,\chi_\rmR)\tilde{\rho},
 \end{align}
 where ${\cal L}(\chi_\rmL,\chi_\rmR)$ is defined implicitly, and
 $\chi_{\rmL/\rmR}$ is the so-called counting field in FCS \cite{Bla001,Naz03}.
 The solution to \Eq{vrho} reads formally
 \begin{align}
 \tilde{\rho}(\chi_\rmL,\chi_\rmR,t)=e^{{\cal L}(\chi_\rmL,\chi_\rmR)(t-t_0)}
 \tilde{\rho}(\chi_\rmL,\chi_\rmR,t_0).
 \end{align}
 We assume that electron counting begins at $t_0$ such that
 $\rho^{(\NL,\NR)}(t_0)=\rho(t_0)\dlt_{\NL,0}\dlt_{\NR,0}$ and
 thus $\tilde{\rho}(\chi_\rmL,\chi_\rmR,t_0)=\rho(t_0)$.
 By performing the inverse Fourier transformation, one readily obtains
 \begin{align}\label{calU}
 \rho^{(\NL,\NR)}(t)&=\int_{-\pi}^{\pi} \frac{\rmd\chi_{\rmL}\rmd\chi_{\rmR}}{(2\pi)^2}
 e^{{\cal L}(\chi_\rmL,\chi_\rmR)(t-t_0)-\rmi\sum_\ell N_\ell \chi_\ell}\rho(t_0)
 \nonumber \\
 &\equiv{\cal U}(\NL,\NR,t-t_0)\rho(t_0),
 \end{align}
 where the implicitly defined ${\cal U}(\NL,\NR,\dlt t)$ actually is
 the number-resolved propagator for the reduced system.
 An important figure of merit of this propagator is that it solely
 depends on the dynamic structure of the quantum master equation,
 rather than the initial reduced quantum state, which makes it
 very efficient in unraveling of the G-QMME into individual quantum
 trajectories.

 Specifically, let us consider the evolution of the reduced state
 during an infinitesimal time interval [$t_k$,$t_k+\rmd t$].
 In general, \Eq{calU} is valid for an arbitrary number of electrons
 ($\NL$ or $\NR$) transmitted trough the SET, depending on the
 time interval ($\rmd t$).
 Yet, here we are interested in the regime where single electron tunneling
 events dominates. We take $\rmd t\ll$ min($\GamL^{-1},\GamR^{-1}$),
 such that the probability of having ($\NL,\NR)\geq2$ is negligible.
 Furthermore, in stead of using $\NL$ and $\NR$ directly, we introduce
 two stochastic point variables $\rmd \NL(t)$ and $\rmd \NR(t)$ (with
 values either 0 or 1) to represent,
 respectively, numbers of electron tunneled through the left and right
 junctions of the SET during the time interval $\rmd t$.
 According to \Eq{calU}, given the condition of a state $\rho(t_k)$ at $t_k$,
 the state at $t_k+\rmd t$ reads
 \be\label{rhoU}
 \rho^{(\rmd \NL,\rmd \NR)}(t_k+\rmd t)
 ={\cal U}(\rmd \NL,\rmd \NR,\rmd t)\rho(t_k),
 \ee
 Here tunneling through the left and right junctions could not
 take place at the same time, due to the sequential tunneling
 picture of the G-QMME (\ref{QME}).
 The stochastic point variables $\rmd \NL(t)$ and $\rmd \NR(t)$
 thus cannot take the value 1 simultaneously.

 If the measurement records are completely ignored (averaged over),
 the ensemble averaged quantum state is given by
 \begin{align}\label{rhoc}
 \rho(t_k+\rmd t)
 =&\sum_{\rmd \NL,\rmd \NR}\rho^{(\rmd \NL,\rmd \NR)}(t_k+\rmd t)
 \nonumber \\
 =&\sum_{\rmd \NL,\rmd \NR}{\rm Pr}(\rmd \NL,\rmd \NR)\rho^{\rmc}(t_k+\rmd t),
 \end{align}
 where ${\rm Pr}(\rmd \NL,\rmd \NR)=\tr\{\rho^{(\rmd \NL,\rmd \NR)}(t_k+\rmd t)\}$
 represents the probability of having $\rmd \NL$ electron tunneled through the left
 junction and $\rmd\NR$ electron through the right one at the time $t_k+\rmd t$,
 and $\tr\{\cdots\}$ denotes the trace over the degrees of freedom of the
 reduced quantum states (qubit plus SET QD).
 $\rho^{\rmc}(t_k+\rmd t)
 =\rho^{(\rmd \NL,\rmd \NR)}(t_k+\rmd t)/{\rm Pr}(\rmd \NL,\rmd \NR)$
 is the normalized state conditioned on the definite measurement
 result at $t_k+\rmd t$.

 In fact, \Eq{dNLR} means that if one generates $\rmd \NL$ and $\rmd \NR$
 stochastically for each time interval $[t_k,t_k+\rmd t]$ and then collapses
 onto a specific state $\rho^{\rmc}(t_k+\rmd t)$ at the end
 of each time interval, one has actually achieved
 a particular single realization of continuous measurements conditioned on
 the specific measurement results.
 The stochastic point variables $\rmd \NL$ and $\rmd \NR$ for single electron
 tunneling events, respectively, satisfy
 \bsube\label{dNLR}
 \begin{gather}
 {\rm E}[\rmd \NL(t)]=\GamL \tr\{d^\dag \rho^{\rmc}(t) d\}\rmd t,
 \\
 {\rm E}[\rmd \NR(t)]=\GamR \tr\{d \rho^{\rmc}(t) d^\dag\}\rmd t,
 \end{gather}
 \esube
 where E[$\cdots$] stands for an ensemble average of a large number of quantum
 trajectories.

 Now, it is apparent that electron tunneling conditions future evolution of
 the reduced state [\Eq{rhoU}], while real-time quantum state conditions
 the observed tunneling events through the left and right junctions [\Eq{dNLR}].
 We have now successfully unraveled the G-QMME into an ensemble of individual
 quantum trajectories.
 Within this unraveling scheme, one is able to propagate the conditioned
 quantum state [$\rho^\rmc(t)$] and the observed result
 [$\rmd N_{\rm L/R}(t)$] in a self-consistent manner.
 Furthermore, the unraveling of the G-QMME enables us to calculate the
 noise power spectrum via the stochastic formalism.
 The fluctuations in the detecting current are characterized
 by the two-time correlation function
 \be\label{curcor}
 G(\tau)=\{{\rm E}[i_\ell(t+\tau)i_\ell(t)]
 -{\rm E}[i_\ell(t+\tau)]{\rm E}[i_\ell(t)]\}|_{t\rightarrow\infty},
 \ee
 where $i_\ell(t)=\rmd N_{\ell}/\rmd t$ is the current through the
 junction $\ell=$\{L, R\} for point process.
 The noise power spectrum of the current is then simply given by
 \be\label{Somg}
 S(\omg)=2\omg \int_{-\infty}^{\infty} \rmd \tau e^{\rmi \omg \tau} G(\tau).
 \ee

 \section{\label{thsec5} Monte Carlo simulation for conditional evolution}

 \begin{figure*}
 \includegraphics*[scale=1.2]{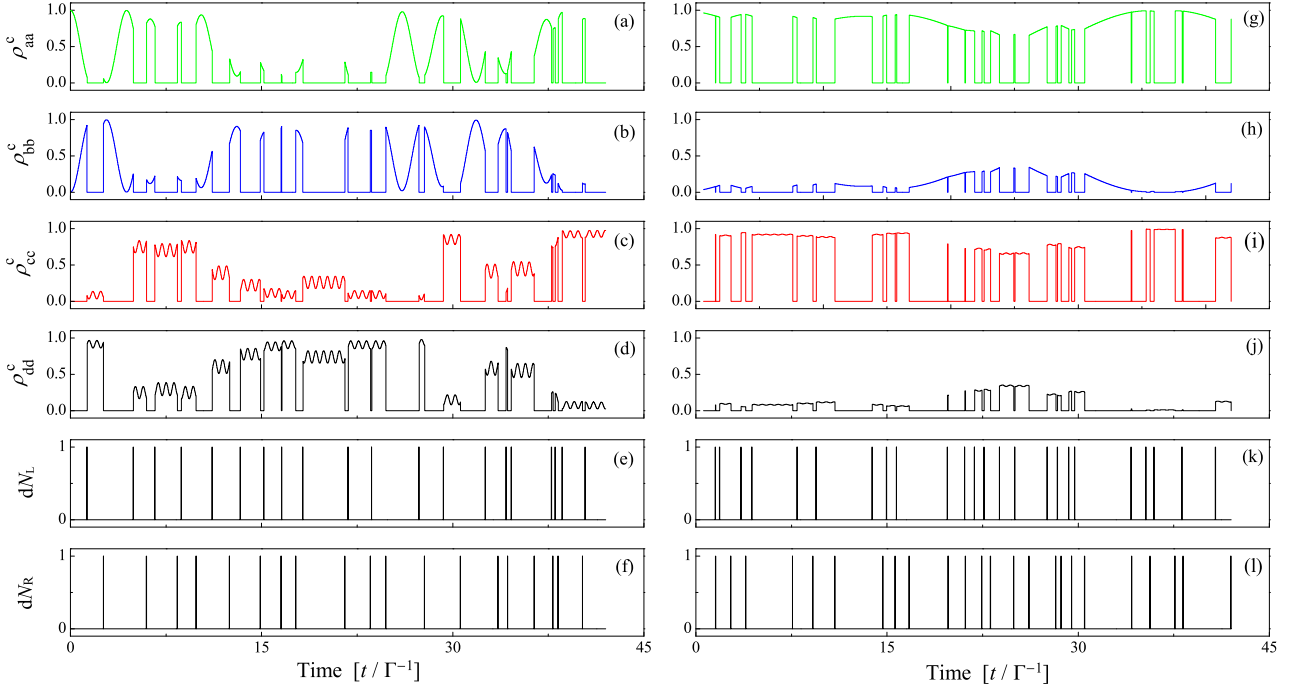}
 \caption{\label{Fig2} Typical quantum trajectories and
 corresponding detection records for $\Omg/\Gam=1.0$ (a)-(f) and
 $\Omg/\Gam=0.1$ (g)-(l), where $\Gam=\GamL+\GamR$ is supposed to
 be a constant, and is used as the unit of energy.
 The initial condition is an empty SET QD with the qubit in
 the logical state $\alf$ as shown in \Fig{Fig1}(a), i.e.,
 $\rho^\rmc(t=0)=|{\rm a}\ra\la {\rm a}|$.
 The measurement voltage is $V/\Gam=5$ ($\mu_{\rm L/R}=\pm 2.5\Gam$),
 such that  $\lambda_+>\mu_\rmL>\lambda_->\mu_\rmR$.
 The involving Fermi functions can be well approximated by either one
 or zero.
 The time step used is $\Delta t$=0.01$\Gam^{-1}$.
 Other plotting parameters are: $\epl=0$, $k_{\rm B} T/\Gam=1.0$, $\GamL=\GamR=\Gam/2$,
 and $\dlt U/\Gam=10$.}
 \end{figure*}

 The possible four electron configurations of the reduced quantum system
 (qubit plus SET) is shown schematically in \Fig{Fig1}.
 The eigenenergies for a symmetric qybit ($\epl=0$) can be readily obtained
 as $\lambda_+\pm\Omg$ and $\lambda_-\pm\Omg$, where
 $\lambda_\pm=\frac{1}{2}(\dlt U\pm \sqrt{\dlt U^2+\Omg^2})$.
 Here, we consider the case of $\dlt U\gg \Omg$. Then the eigenenergies
 may be greatly simplified,
 i.e. $\lambda_+\pm\Omg\approx\lambda_+$ and $\lambda_-\pm\Omg\approx \lambda_-$.
 Under the application of an appropriate bias voltage
 $\lambda_+>\mu_\rmL>\lambda_->\mu_\rmR$,
 the involving Fermi functions in the tunneling rates are
 approximated by either one or zero, and the resultant
 quantum master equation describing the reduced dynamics can be remarkably
 simplified. Let us denote the density matrix elements by $\rho_{jj'}$
 where $j,j'=\{{\rm a, b, c, d}\}$ corresponds to one of the electron
 configurations as shown in \Fig{Fig1}(a)-(d), respectively.
 The $\chi$-resolved quantum master equation (\ref{vrho}) reads
 \bsube\label{QMEele}
 \begin{align}
 \frac{\partial}{\partial t}\tilde{\rho}_{\rm aa}
 =&\rmi\Omg(\tilde{\rho}_{\rm ab}-\tilde{\rho}_{\rm ba})-\GamL\tilde{\rho}_{\rm aa}
 +e^{\rmi \chi_\rmR}\GamR\tilde{\rho}_{\rm cc},
 \\
 \frac{\partial}{\partial t}\tilde{\rho}_{\rm bb}
 =&\rmi\Omg(\tilde{\rho}_{\rm ba}-\tilde{\rho}_{\rm ab})
 +(e^{\rmi \chi_\rmL}\GamL+e^{\rmi \chi_\rmR}\GamR)\tilde{\rho}_{\rm dd},
 \\
 \frac{\partial}{\partial t}\tilde{\rho}_{\rm cc}
 =&\rmi\Omg(\tilde{\rho}_{\rm cd}-\tilde{\rho}_{\rm dc})
 +e^{-\rmi \chi_\rmL}\GamL\tilde{\rho}_{\rm aa}-\GamR\tilde{\rho}_{\rm cc},
 \\
 \frac{\partial}{\partial t}\tilde{\rho}_{\rm dd}
 =&\rmi\Omg(\tilde{\rho}_{\rm dc}-\tilde{\rho}_{\rm cd})-(\GamL+\GamR)\tilde{\rho}_{\rm dd},
 \\
 \frac{\partial}{\partial t}\tilde{\rho}_{\rm ab}
 =&\rmi\Omg(\tilde{\rho}_{\rm aa}-\tilde{\rho}_{\rm bb})+\rmi\Lmd (\tilde{\rho}_{\rm ab}-\tilde{\rho}_{\rm cd})
 -\frac{1}{2}\GamL\tilde{\rho}_{\rm ab}
 \nl
 & +\left(\frac{1}{2}e^{\rmi \chi_\rmL}\GamL+e^{\rmi \chi_\rmR}\GamR\right)\tilde{\rho}_{\rm cd},\label{varrhoab}
 \\
 \frac{\partial}{\partial t}\tilde{\rho}_{\rm cd}
 =&\rmi\Omg(\tilde{\rho}_{\rm cc}-\tilde{\rho}_{\rm dd})-\rmi \Lmd (\tilde{\rho}_{\rm ab}-\tilde{\rho}_{\rm cd})
 +\frac{1}{2}e^{-\rmi \chi_\rmL}\GamL\tilde{\rho}_{\rm ab}
 \nl
 &+\left(-\frac{1}{2}\GamL-\GamR+\rmi\dlt U\right)\tilde{\rho}_{\rm cd},\label{varrhocd}
 \end{align}
 \esube
 where $\Lmd$, the renormalization of qubit level mismatch arising from
 bath spectral functions [cf. \Eq{disp}], is given by \cite{Luo104904}
 \be\label{Lambda}
 \Lmd=\sum_{\ell=\rmL,\rmR}\sum_{\pm}\frac{\Gam_\ell}{2\pi}{\rm Re}\left[
 \Psi\left(\frac{1}{2}-\rmi \frac{\lambda_\pm-\mu_\ell}{2\pi k_{\rm B} T}
 \right)\right].
 \ee
 Here $\Psi$ is the digamma function, $\mu_\ell $ is the chemical potential
 of left ($\ell=$L) or right ($\ell$=R) SET electrode, $k_{\rm B}$ is
 the Boltzmann constant, and $T$ is the temperature.
 It has been revealed that this renormalization has an important impact
 on the signal-to-noise ratio in continuous measurement of a chage qubit.
 Apparently, this effect is absent in the Lindblad master equation.

 \begin{figure*}
 \includegraphics*[scale=0.9]{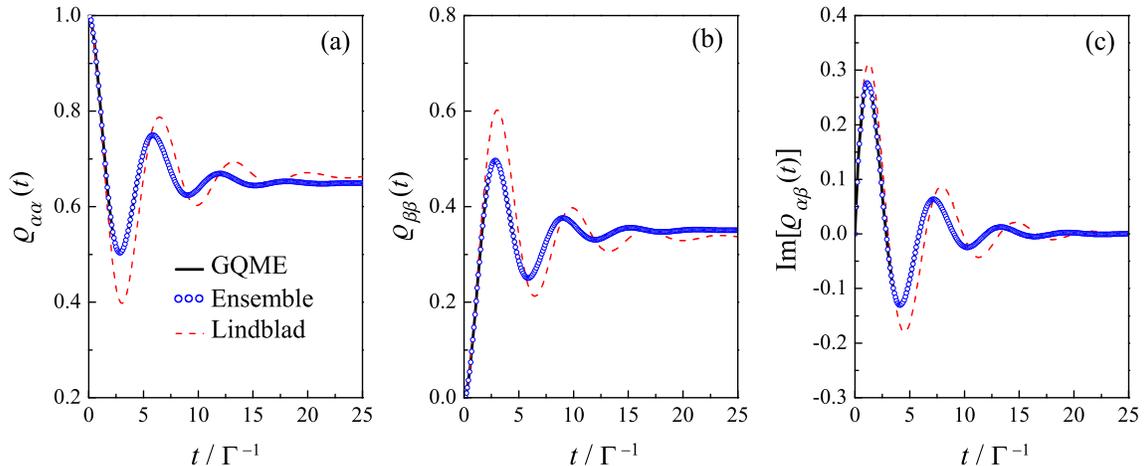}
 \caption{\label{Fig3}
 Unconditional measurement dynamics obtained by using the G-QMME
 (\ref{QME}) (solid curves), ensemble average over 20 thousand quantum
 trajectories similar to those in \Fig{Fig2} (symbols).
 For comparison, the results using Lindblad master equation are
 also plotted by the dashed curves.
 The initial condition is $\rho(t=0)=|\alpha\ra\la\alpha|$.
 The total tunneling width $\Gam=\GamL+\GamR$ is kept constant,
 and used as the unit of energy.
 Other plotting parameters used are: $\epl=0$, $\Omg/\Gam=0.5$, $V/\Gam=5$,
 $k_{\rm B}T/\Gam=1$, $\GamL=\GamR=\Gam/2$, and $\dlt U/\Gamma=10$.}
 \end{figure*}

 We are now in a position to employ the theory developed in section \ref{thsec4}
 to unravel the quantum master equation (\ref{QMEele}) into individual
 quantum trajectories, which will be used to implement feedback control
 of the charge qubit later.
 The numerical results for conditional state evolution and corresponding
 tunneling events are displayed in \Fig{Fig2} for different values
 of qubit interdot coupling $\Omg$.
 The initial condition corresponds to the charge configuration as shown in
 \Fig{Fig1}(a), i.e. $\rho^\rmc(t=0)=|{\rm a}\ra\la {\rm a}|$.
 For $\Omg/\Gam=1.0$, the qubit exhibits oscillations between the quantum
 states $|{\rm a}\ra$ and $|{\rm b}\ra$, which is stochastically interrupted
 whenever one electron tunnels into the SET [see \Fig{Fig2}(e) for ${\rm d}N_\rmL$].
 The electron may dwell on the SET for a random amount of time,
 where it experiences fast oscillations with frequency $\sim\lambda_+$,
 before it tunnels out of the SET QD.
 The conditional state evolution and corresponding tunneling events become
 quite different in the case of suppressed interdot coupling $\Omg/\Gam=0.1$.
 One observes very slow oscillations between $|{\rm a}\ra$ and $|{\rm b}\ra$,
 as shown in \Fig{Fig2}(g)-(h).
 Unambiguously, we find bunching of electron tunneling
 events through the SET; see \Fig{Fig2}(k)-(l).
 When the qubit stays in the logical state $|\beta\ra$, it will
 prevent electrons tunnel through the SET due to the strong Coulomb
 repulsion between the qubit and SET QD ($\dlt U/\Gam\gg1$).
 Electrons can only flow in short time windows where the qubit
 relaxes to the state $|\alf\ra$, leading eventually to the bunching
 of tunneling events.

 The occurrence of the bunching of tunneling events is also manifested
 in the FCS of SET current. By using the number-resolved quantum master
 equation (\ref{CQME}), the first cumulant (current) is given by
 \be
 \la I \ra=\frac{\GamL \GamR}{\GamL +2\GamR},
 \ee
 which is apparently independent of the dynamical structure of the
 qubit.
 The second cumulant $\la I^2 \ra$ is associated to the shot noise.
 The corresponding Fano factor $F\equiv \la I^2 \ra/\la I\ra$ is given by
 \be
 F=\frac{\GamL^2+4\GamR^2}{(\GamL+2\GamR)^2}
 +\frac{\GamL^2 \GamR^2}{2(\GamL+2\GamR)^2 \Omega^2}
 +{\cal O}\left(\frac{1}{\dlt U}\right),
 \ee
 which unambiguously depends on the the qubit parameter ($\Omg$).
 Here, the first term  is simply the shot noise of a double-barrier
 system \cite{Bla001,Che914534,Luo07085325} and is definitely below the Poisson value.
 The influence of qubit dynamics on the SET transport is characterized
 by the second term.
 The noise may be remarkably enhanced as the qubit interdot coupling
 $\Omg$ decreases, leading to bunching of tunneling events as shown
 in \Fig{Fig2}(k)-(l).
 Our result thus shows unambiguously that the intrinsic dynamics of
 the qubit may serve as an essential mechanism that leads to bunching
 of tunneling events, and finally to a strong super-Poissonian noise
 in transport through a single QD SET.

 \begin{figure*}
 \includegraphics*[scale=1]{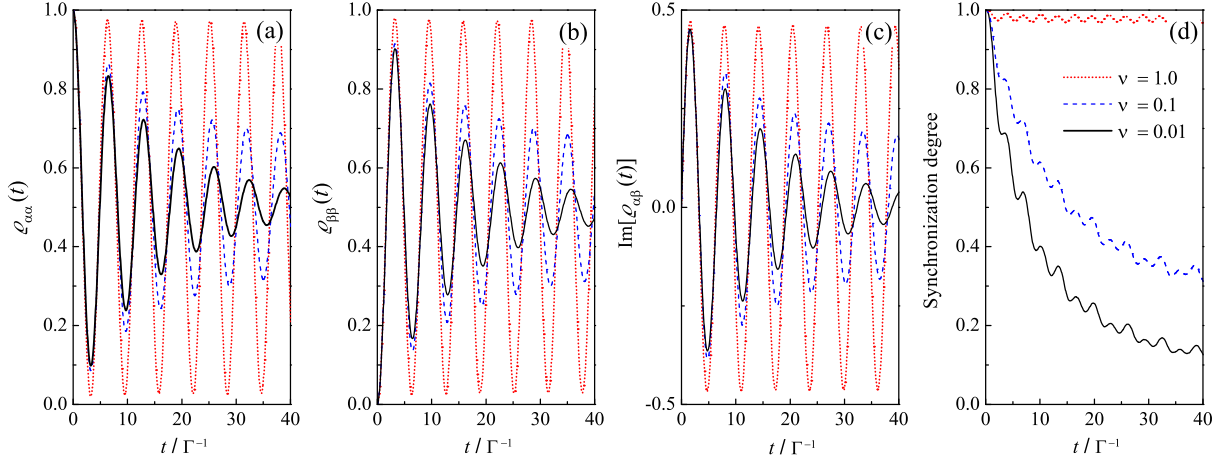}
 \caption{\label{Fig4}
 The dynamics of the charge qubit and corresponding degree of synchronization
 for different values of feedback strength $\nu=0.01$ (solid curves), $\nu=0.1$
 (dashed curves), and $\nu=1.0$ (dotted curves).
 The results are obtained by ensemble average over 20 thousand
 quantum trajectories.
 The total tunneling width $\Gam=\GamL+\GamR$ is constant and used as the unit
 of energy.
 Other plotting parameters used are the same as those in \Fig{Fig3}.}
 \end{figure*}

 To further verify the validity of our Monte Carlo method for
 conditional evolution of the reduced state, it is instructive
 to make ensemble average over a large number of
 quantum trajectories analogous to those in \Fig{Fig2}, and then
 compare with the results obtained using the G-QMME (\ref{QME}).
 Yet, the G-QMME or the quantum trajectory describes the propagation
 of the combined (qubit plus SET QD) system.
 One thus has to trace over the degrees of freedom of the QD to
 obtain the dynamics of the qubit alone
 \be\label{trSET}
 \varrho(t)=\tr_{\rm SET}\{\rho(t)\},
 \ee
 where ${\rm tr}_{\rm SET}\{\cdots\}$ stands for the trace over the degrees
 of freedom of the SET QD.
 One may expect perform the trace directly upon \Eq{QME}; Yet, a closed form
 of the master equation for the qubit alone cannot be obtained without
 further approximations.
 One possible method is to assume extremely asymmetric tunnel
 junctions for the SET and then adiabatically eliminate the degrees
 of freedom of the SET island to arrive at the reduced density
 matrix for the qubit alone \cite{Wis01235308}.
 Actually, such an assumption is equivalent to treating the SET effectively
 as a single junction detector, analogous to a quantum point contact.

 Here, we first numerically propagate the combined (qubit plus SET)
 system, and then trace over the degrees of freedom of the SET QD,
 i.e. \Eq{trSET} to get the dynamics of the qubit alone.
 The numerical result is shown in \Fig{Fig3}.
 The reduced dynamics by using G-QMME (solid curves) and by
 ensemble average over 20 thousand quantum trajectories (symbols)
 show striking agreement, and thus verifies the validity of
 the unraveling scheme proposed in section \ref{thsec4}.
 For comparison, the result obtained using the Lindblad master
 equation is also plotted by the dashed curves in \Fig{Fig3}.
 Notable difference is observed, which arises actually from
 the renormalization of the level mismatch $\Lambda$ in
 \Eq{Lambda}.
%

 \section{\label{thsec6}Implementation of Feedback control}

 So far, we have presented a Monte Carlo method to simulate the continuous
 quantum measurement of a charge qubit by an SET detector.
 Now we are in a position to implement the feedback control of coherent
 evolution of the charge qubit by utilizing the unraveling of the underlying
 measurement dynamics governed by the G-QMME, which is not necessarily of
 non-Lindblad form.
 In the dot state of the qubit ($|\alf\ra$ and $|\beta\ra$), the desired
 (target) pure state under protect is
 \be\label{psid}
 |\psi_\rmd(t)\ra=\cos(\Omg t)|\alf\ra+\rmi \sin(\Omg t)|\beta\ra
 \ee
 where $\Omg$ is the inherent qubit interdot coupling.

 The basic idea of the feedback is to convert the detector's output
 into the evolution of a qubit state $\varrho_\rmc(t)$, with
 $\varrho^\rmc(t)=\tr_{\rm SET}\{\rho_\rmc(t)\}$.
 The real-time state $\varrho^\rmc(t)$ is then compared with the desired
 state $\varrho_\rmd (t)\equiv |\psi_\rmd(t)\ra\la\psi_\rmd(t)|$, their 
 difference is utilized to design the
 feedback Hamiltonian such that their difference can be reduced in the
 next evolution step.
 In each successive step, the feedback Hamiltonian acts only for
 an infinitesimal time interval $\rmd t$.
 According to the suboptimal algorithm \cite{Doh01062306}, state
 propagation in each infinitesimal time step maximizes the fidelity
 of $\varrho^\rmc(t)$ with $\varrho_\rmd(t)$.
 Specifically, let us consider state evolution concerning the feedback
 Hamiltonian, the state $\varrho^\rmc(t+\rmd t)$ is given by
 \begin{align}
 \varrho^\rmc(t+\rmd t)=&\varrho^\rmc(t)-\rmi [H_{\rm fb},\varrho^{\rmc}(t)]\rmd t
 \nonumber \\
 &-\frac{1}{2}[H_{\rm fb},[H_{\rm fb},\varrho^\rmc(t)]](\rmd t)^2+\cdots,
 \end{align}
 where the feedback Hamiltonian $H_{\rm fb}$ is to be determined via
 corresponding restrictions.

 \begin{figure*}
 \includegraphics*[scale=1.1]{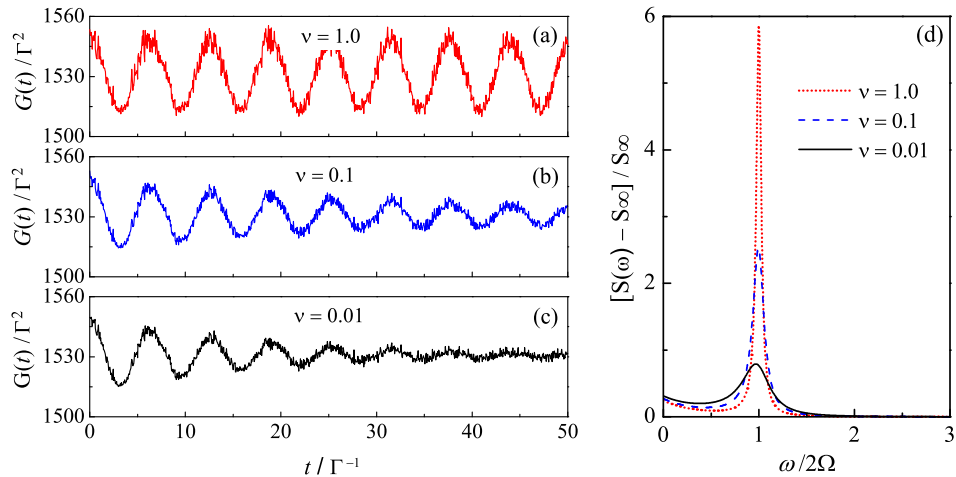}
 \caption{\label{Fig5} Correlation functions of transport current through
 the left junction and the corresponding noise spectrum for different values
 of feedback strength $\nu=0.01$ (solid curves), $\nu=0.1$ (dashed curves),
 and $\nu=1.0$ (dotted curves).
 The correlations are obtained by ensemble average over 8,000 quantum
 trajectories.
 The total tunneling width $\Gam=\GamL+\GamR$ is set as the unit of energy.
 Other plotting parameters are the same as those in \Fig{Fig3}.}
 \end{figure*}

 The fidelity of the state $\varrho^\rmc(t+\rmd t)$ with the target state
 then is simply given by
 \begin{align}
 {\rm Fid}\equiv&\la \psi_\rmd|\varrho^\rmc(t+\rmd t)|\psi_\rmd\ra
 \nonumber \\
 =&\la \psi_\rmd|\varrho^\rmc(t)|\psi_\rmd\ra
 -\rmi\la \psi_\rmd|[H_{\rm fd},\varrho^\rmc(t)]|\psi_\rmd\ra\rmd t
 \nonumber \\
 &-{\textstyle \frac{1}{2}}
 \la \psi_\rmd|[H_{\rm fb},[H_{\rm fb},\varrho^\rmc(t)]]|\psi_\rmd\ra(\rmd t)^2+\cdots.
 \end{align}
 To optimize the fidelity, one should maximize the dominant term, i.e. the one
 proportional to $\rmd t$.
 This imposes certain constraints, e.g.
 the sum of the squares of the eigenvalues,
 the sum of the norms of the eigenvalues,
 or the restriction on the maximum eigenvalue of $H_{\rm fb}$, etc.
 Actually, these constraints originate from the limitation on
 the feedback strength or finite Hamiltonian resources.
 By employing the first type of constraint, i.e.
 $\tr_{\rm qb}\{H^2_{\rm fb}\}\leq \nu$ with $\tr_{\rm qb}\{\cdots\}$
 the trace over the qubit states, the feedback Hamiltonian
 is constructed as \cite{Wan07155304,Jin06233302}
 \be\label{Hfb}
 H_{\rm fb}=\rmi \kappa[|\psi_\rmd(t)\ra\la \psi_\rmd(t)|,\varrho^{\rmc}(t)],
 \ee
 where $\kappa=\sqrt{\frac{\nu}{2(p-q)}}$, with
 $p=\la \psi_\rmd|[\varrho^\rmc(t)]^2|\psi_\rmd\ra$ and
 $q=[\la \psi_\rmd|\varrho^\rmc(t)|\psi_\rmd\ra]^2$.

 To make it much more easily accessible for experimentalists
 to implement the above feedback control protocol, we insert \Eq{psid}
 into \Eq{Hfb}, and eventually arrive at a translation
 of the feedback Hamiltonian as
 \be\label{Hfb2}
 H_{\rm fb}=K\sigma_x
 \ee
 with
 \be
 K=\left\{
   \begin{array}{ll}
    +\sqrt{\frac{\nu}{2}}, & \Dlt \phi<0 \\
    -\sqrt{\frac{\nu}{2}}, & \Dlt \phi>0
   \end{array}.
 \right.
 \ee
 It resembles a simple {\it bang-bang} control, where the feedback
 parameter between two values is simply determined by the phase
 ``error'' defined as $\Dlt\phi\equiv\phi(t)-\phi_0$,
 with $\phi(t)= \arctan(2{\rm Im}\{\varrho^\rmc_{\alf\beta}(t)\}
 /[\varrho^\rmc_{\alf\alf}(t)-\varrho^\rmc_{\beta\beta}(t)])$ the
 relative phase between the logical states ``$|\alf\ra$''
 and ``$|\beta\ra$'',
 and $\phi_0=2\Omg t$(mod $2\pi$).
 It is worthwhile to mention that the two versions of the feedback
 Hamiltonian protocols, i.e. \Eq{Hfb} and \Eq{Hfb2} are totally
 equivalent to each. Yet, second one actually provided a much easier
 way for experimentalists to implement.

 In \Fig{Fig4}, the effect of feedback control is plotted for various
 values of feedback strength $\nu$.
 The feedback is implemented into every single quantum trajectories,
 analogous to those in \Fig{Fig2}.
 Eventually the propagation of the state is obtained by an ensemble
 average over 20 thousand trajectories.
 For sufficient large feedback strength (see the dotted curves for $\nu$=1),
 coherent oscillation of the charge quit can be maintained, in principle,
 for an arbitrarily long time.
 To quantitatively characterize how close to the desired state
 the protected state reach, we introduce the synchronization degree,
 defined as
 $D\equiv2\tr_{\rm qb}\{\varrho^\rmc\varrho_\rmd\}-1$.
 For $D = 1$, it means complete synchronization, indicating that
 the state is perfectly protected via the feedback control.
 The results for various feedback strengths are shown in \Fig{Fig4}(d).
 Apparently, for a large feedback strength ($\nu=1$), the synchronization
 degree can reach almost the maximum value of 1, showing thus the
 high effectiveness of our feedback scheme.

 Another important figure of merit of the coherent oscillations
 is the signal-to-noise ratio of the detector's output.
 To investigate this important feature of feedback effect, we first
 evaluate the correlation function of
 current transport through the left junction of the SET (the right one
 gives similar results) according to \Eq{curcor}.
 The numerical results obtained by ensemble average over 8,000 quantum
 trajectories are displayed in \Fig{Fig5} for different
 feedback strength $\nu$.
 As $\nu$ increases, the correlation function demonstrates clearly
 coherent oscillation behavior; see \Fig{Fig5}(a) for $\nu=1.0$.
 With the knowledge of the current correlation function, the corresponding
 noise spectrum can be obtained directly by performing the Fourier
 transformation, as shown in \Eq{Somg}.
 The numerical results are shown in \Fig{Fig5}(d).
 An enhancement of the feedback strength leads to a rising and sharping
 of the coherent peak located at frequency $\omg=2\Omg$.
 Actually, the height of this peak relative to its pedestal is a measure
 of the signa-to-noise ratio.
 It was argued in that the signal-to-noise ratio is limited
 at 4, known as the Korotkov-Averin bound \cite{Kor01085312,Kor01165310}.
 It was revealed the signal-to-noise ratio of an SET detector can not
 reach the limit of an ideal detector \cite{Gur05073303}.
 However, we find the signal-to-noise ratio can be remarkably increased
 in the case of strong feedback strength (see the dotted curve),
 even violating Korotkov-Averin bound.
 Our feedback protocol thus serves an effective method to improve
 the signal-to-noise ratio that can exceed the up bound limit of 4,
 indicating ideal qubit coherent oscillations under continuous
 measurement by an SET detector.

 Recently, a closed-loop feedback scheme was proposed for the
 stabilization of a pure qubit state by employing, analogously,
 a capacitively coupled SET.
 There, the real-time tunneling events through SET detector was
 directly back-coupled into qubit parameters.
 This purification process was found to be independent of the initial
 state and could be accomplished simply after a few electron jumps
 through the SET detector.
 The qubit states is stabilized above a critical detector-qubit
 coupling.
 In comparison, the advantage of our feedback scheme is that a
 coherent oscillations of the qubit could be maintained for arbitrary
 qubit parameters, as long as the feedback strength is sufficient strong.

\section{\label{thsec7}Summary}

 In the context of a charge qubit under continuous monitoring
 by a single electron transistor, we have proposed an unraveling
 scheme of the G-QMME into an ensemble
 of individual quantum trajectories specifically for stochastic
 point process.
 A suboptimal feedback algorism is implemented into every single
 quantum trajectory to protect a desired pure state.
 It is demonstrated that coherent oscillation of the charge qubit
 can be maintained for an arbitrarily long time in the case of
 sufficient feedback strength.
 The corresponding synchronization degree can reach almost close to
 the maximum value of 1.
 It is also observed that the signal-to-noise ratio increases
 with rising feedback strength, and remarkably could even exceed
 the well-known Korotkov-Averin bound in quantum measurement, reflecting
 actually ideal quantum coherent oscillations of the qubit.
 The proposed unraveling and feedback scheme thus may serve as an
 essential mechanism to sustain ideal quantum coherent oscillations
 in a very transparent and straightforward manner.
 It is also highly expect that it may have important applications
 in the field of solid-state in quantum computational algorithms.

 \begin{acknowledgments}
 Support from the National Natural Science Foundation of
 China (11204272 and 11274085) and the Natural Science Foundation of
 Zhejiang Province (LZ13A040002 and Y6110467) are gratefully acknowledged.
 \end{acknowledgments}


\end{document}